\documentclass[a4paper,11pt]{article}
\usepackage{jheppub} 
\usepackage{lineno}
\usepackage{amsmath}
\usepackage{tikz}
\usepackage{spectralsequences}
\usepackage{hyperref}
\usepackage{jheppub}

\usetikzlibrary{matrix}
\newcommand*\Z{\mathbb{Z}}
\newcommand*\ZZ{|[draw,color=red,rectangle]| \Z}

\title{\boldmath Anomaly Matching in 6d $\mathcal{N}=(2,0)$ SCFTs from M5 Cobordism}

\author{Aiden Sheckler}
\affiliation{Department of Physics, University of California, San Diego,\\
9500 Gilman Drive, La Jolla CA 92093-0319, USA}

\emailAdd{asheckler@ucsd.edu}

\abstract{We investigate the anomalies of 6d $\mathcal{N}=(2,0)$ superconformal field theories for any ADE gauge group using the modern characterization of anomalies by cobordism. We propose that, in order to account for all features of the anomaly, a bordism theory with exotic tangential structure is needed. We then attempt to match the anomaly in the IR effective theory on the tensor branch with a suitable WZW term. Once again, we show that the exotic bordism structure is necessary to achieve this. Our results suggest a natural explanation for the origin of the Hopf-Wess-Zumino term. }

\begin{document}
\maketitle
\flushbottom

\section{Introduction}
\label{sec:intro}

A fundamental property of a symmetry in quantum field theory is its ability to possess a 't Hooft anomaly. The anomaly manifests as an obstruction to gauging the symmetry. Parallel to the recent evolution in our understanding of symmetries, there has advanced an investigation in to the origin, classification, and consequences of these anomalies. Among these developments have come new proposals for the classification of anomalies by inflow. These are topological quantum field theories which take the form of topological invariants defined by generalized cohomology \cite{Freed_2021, Kapustin_2015, Yonekura_2019, Gaiotto_2019, Xiong_2018, Freed_2006}. Our purpose in this work is to bring to bear some of these recent insights in order to study anomalies belonging to a class of theories which excites intense interest in quantum field theory and string theory. These are the 6d $\mathcal{N}=(2,0)$ superconformal field theories. \\

The study of these theories has been immensely rewarding, and they play a central role in the study of gauge theories and conformal field theories (see \cite{Moore_FK, Heckman:2018jxk} for a review). They are conformal field theories with the maximum possible superconformal symmetry in the maximum possible dimension. A rich trove of insights has been recovered from their investigation \cite{Gaiotto:2009we,  Alday:2009aq, Witten:2009at, Witten:1996hc}. Moreover the 6d (2,0) theory gives rise to a vast landscape of lower-dimensional theories under compactification. This perspective has allowed for deep insights into strongly-coupled phases of gauge theories, dualities, and an expansive network of relations amongst the landscape. \\

Some of the greatest motivation for the study of these theories comes from the many enduring mysteries surrounding them. They have no known Lagrangian description, and no semiclassical limit in which we can perform perturbation theory. However top-down constructions from string theory are available, via brane constructions. Many results have also been obtained by conformal bootstrap methods \cite{Beem:2015aoa}. While the superconformal theory is obtained in the limit of tensionless strings, it is believed to describe a local quantum field theory \cite{Seiberg_1997, Seiberg_1996}. \\

The unorthodox constructions of these theories makes their study challenging. The anomalies of these theories therefore provide precious data about the strongly-coupled regime. The anomalies of the 6d theory are studied by determining the 7d topological anomaly theory which matches the anomaly by inflow. This is generally a relative theory \cite{freed2014relativequantumfieldtheory, Gukov:2020btk} which gives rise to a space of conformal blocks, so that the partition function of the 6d theory should really be thought of as a ``partition vector". The local form of these anomalies has been known for many years \cite{Harvey_1998,Intriligator_2000,Ohmori:2014kda}, while the full global anomalies were computed for the A-type theories and conjectured for the D, E theories more recently \cite{monnier2017anomalyfieldtheoriessixdimensional, Monnier_2014}. \\

The anomalies have many interesting topological contributions. In particular, there exists some exotic data which is argued to be necessary to define the 6d (2,0) theory \cite{monnier2017anomalyfieldtheoriessixdimensional}. A key observation in our discussion is that this data plays a central role in the structure of the anomaly. This data includes a rank-5 $R$-symmetry bundle, a choice of Euler structure, and a restriction of the M-theory C-field. We review this data as well as its contributions to the A-type anomalies in \hyperref[sec2]{Section 2}. \\

The manifestation of this data in the anomaly theory naturally raises the question of the classification of these anomalies by a bulk invertible field theory. Invertible field theories are defined as topological invariants associated to some generalized cohomology theory. In particular, for anomalies which depend critically on some choice of tangential structure like spin structure, this generalized cohomology theory is taken to be the associated cobordism theory which is Anderson dual to the bordism which preserves this tangential structure. For example, many anomalies in theories of fermions are defined as spin-bordism invariants. \\

In the course of this paper we will attempt to incorporate these exotic anomaly theories into the modern classification of anomaly theories by generalized cohomology. We will achieve this using a bordism theory $\Omega^{M5}_{\bullet} $ known as \textit{M5 bordism}, which incorporates all of the 6d (2,0) data that appears in the anomaly theory, and was proposed in \cite{monnier2017anomalyfieldtheoriessixdimensional, Monnier_2015}. We propose that the 7d anomaly theories for the anomaly theories of the 6d (2,0) SCFTs are then M5 bordism invariants, and thus correspond to elements of the 8th Anderson-dual cobordism groups $\text{Inv}_{M5}^7=(D\Omega^{M5})^8$. \\

However the main focus of this work is \textit{anomaly matching}. Perhaps the most useful application of 't Hooft anomalies is that they are invariant across renormalization group flows, and thus must be matched between the high and low energy regimes of a theory. The matching of the local anomalies for 6d SCFTs on the tensor branch has been studied extensively \cite{Intriligator_2000, Harvey_1998, Intriligator_2014, Ohmori_2014, heckman2015geometry6drgflows, C_rdova_2016,Cordova:2015vwa}. In particular, for the (2,0) theory the tensor branch of the moduli space of supersymmetric vacua is described by $(\mathbb{R}^5)^{r(G)}/W_G$ where $r(G)$ is the rank of the ADE group and $W_G$ is the Weyl group. The $SO(5)_R$ symmetry is broken away from the superconformal origin when we give the moduli parameters a nonzero vev. In the M5 brane construction of the theory, this physically corresponds to separating the branes along a direction specified by the moduli parameters. The low-energy effective theory can then be studied far out on the tensor branch along directions where the $R$-symmetry is spontaneously broken to $SO(5)_R\to SO(4)_R$. The massless degress of freedom then include a $\mathbb{S}^4$ $\sigma$-model. \\

The anomaly matching for this SSB was studied \cite{Intriligator_2000}, and it was found that a WZW term was necessary to match the local anomaly in the IR. This term was dubbed the \textit{Hopf-Wess-Zumino term}, since it takes the local form of the Hopf invariant for the quaternionic Hopf fibration $\mathbb{S}^3\to \mathbb{S}^7\to \mathbb{S}^4$. This term has been well-studied since \cite{Fiorenza_2021,Kalkkinen_2003, Hu_2011, arvanitakis2019branewesszuminotermsaksz}, and we review it in \hyperref[sec23]{Section 2.3}. However the question of its classification has been noticed since it was first discovered. Indeed, WZW terms are traditionally defined by transgression as cohomology classes of the target space. However clearly in this case $H^7(\mathbb{S}^4)=0$, so such a characterization does not apply. Moreover, there exists a more modern diagnosis of WZW terms by cobordism groups \cite{Lee_2021, freed2007pionsgeneralizedcohomology, Freed_2018, saito2024wesszuminowittentermsspqcd, lee2021comments6dglobalgauge}, but this classification also fails for the usual choices of cobordism, as we discuss in \hyperref[sec42]{Section 4.2}. \\

The primary goal of this work is to resolve this question. We study the anomaly matching of the 6d (2,0) theory using our proposal that the anomalies are classified by M5 cobordism. From this perspective, we will find a natural candidate for the origin of the Hopf-Wess-Zumino term. We achieve this by proposing a generalization of the usual transgression construction of WZW terms, which aligns with the conventional approach of anomaly matching in low-energy effective theories.  \\ 

Other proposals for the characterization of the Hopf-Wess-Zumino term have also been put forth \cite{Fiorenza_2021}, which attempt to describe it using the hypothesis that M5 branes are charge quantized by twisted cohomotopy. In this work we attempt to formulate a classification which aligns with the traditional definition of WZW terms by transgression. It would be interesting to further understand the connection between these proposals and our results we derive here. \\

The plan for this paper is as follows: in \hyperref[sec2]{Section 2} we review background on the 6d (2,0) theories and the known aspects of their anomalies. In particular, we review the data which is needed to define these theories, as well as some motivation for it. We then review the 7d anomaly theory for the A-type theories as derived in \cite{monnier2017anomalyfieldtheoriessixdimensional}. We also review the anomaly matching on the tensor branch and the role of the Hopf-Wess-Zumino term. In \hyperref[sec3]{Section 3} we propose a classification for the anomaly by M5 cobordism. We construct this bordism theory and perform the computations for the invertible phases up to dimension 7, and then compare with the known form of the A-type anomalies. Finally in \hyperref[sec4]{Section 4} we discuss the anomaly matching, beginning with a review of how WZW terms are traditionally defined, as well as their more modern characterization in terms of invertible phases. We then illustrate how these methods fail when applied to the case of the Hopf-Wess-Zumino term. Finally, we generalize these constructions in order to use the proposed M5 bordism characterization of the anomaly, and calculate the possible WZW terms up to dimension 6. The relevant bordism and spectral sequence calculations are left to the \hyperref[appa]{Appendix}.

\newpage

\section{Anomalies of 6d $\mathcal{N}=(2,0)$ Theories}
\label{sec2}
 
We now review some known aspects of the anomaly theory for the 6d (2,0) theory. We first review the results of \cite{monnier2017anomalyfieldtheoriessixdimensional, Monnier_2014}, which derive explicit 7d field theories defining the anomaly. We begin by recounting the relevant data required to define the (2,0) SCFT, which are crucial to encapsulating all features of the anomaly field theory. We then describe the structure and role of each contributing factor to the 7d anomaly field theory. Finally, we review known results about the matching of the anomaly by the Hopf-Wess-Zumino term, far out on the tensor branch of the moduli space.  \\

\subsection{Data of 6d $\mathcal{N}=(2,0)$ Theories}
\label{sec21}

The 6d $\mathcal{N}=(2,0)$ superconformal field theories have no known Lagrangian description, but nonetheless there are several avenues of approach that can be used to study them. The $A_N$ series of 6d (2,0) theories with $\frak{g}= SU(N)$ gauge group can be obtained from type IIA string theory by considering the zero string coupling limit $g_s\to 0$ of a stack of $N$ coincident NS5 branes \cite{Strominger_1996}. This can equivalently be realized as a stack of M5 branes in M-theory \cite{Strominger:1995ac}. From this perspective the $D_N$ series with gauge group $SO(N)$ can be obtained from a stack of orientifold M5 branes, but there is no such known construction for the $E$-type theories. Applying T-duality transverse to the stack produces type IIB string theory on an $A_{N-1}$ singularity. This perspective allows for more general constructions of the 6d (2,0) theory for any ADE Lie algebra $\frak{g}$, by realizing IIB on an ALE singularity $\mathbb{C}^2/\Gamma$ for finite $\Gamma\subset SU(2)$ corresponding to the appropriate Lie algebra. A unifying perspective on such constructions is provided by F-theory, as reviewed in \cite{Heckman:2018jxk}. \\

We briefly recall the pieces of data which are thought to be necessary in order to define the 6d (2,0) SCFT for a given choice of ADE algebra from such constructions. More thorough review and justification can be found in \cite{monnier2017anomalyfieldtheoriessixdimensional, Monnier_2014, Monnier_2015, Monnier_2017_2,Monnier_2015_2,Witten_1997, Witten_2000} and references therein. The data which must be specified is:

\begin{enumerate}
    \item The appropriate framing data on the spacetime manifold $M$, including a choice of orientation, smooth structure, and metric. Notice that in general we do \textit{not} require a spin structure.
    
    \item A rank-5 vector bundle $N$ for the $R$-symmetry $SO(5)_R$, along with specified connection and orientation. In the M5 brane construction of the 6d (2,0) theory this bundle describes the normal bundle parametrizing the transverse directions of the M5 brane stack worldvolume in the ambient M-theory spacetime. 

    \item A generalized spin structure on $TM \oplus N$, which amounts to a trivialization of $w_2(TM)+w_2(N)$. This is necessary to define the fermionic fields in the tensor multiplet on the Coulomb branch. 
    
    \item A restriction of the M-theory C-field, specified by a degree-4 shifted differential cocycle $\check C_M$. 

    \item An Euler structure on the $R$-symmetry bundle $N$.

\end{enumerate}

The last two pieces of data are the most subtle, and require further comment \cite{monnier2017anomalyfieldtheoriessixdimensional, Monnier_2014, Monnier_2015}. They can most clearly be seen from the M5 brane construction of the 6d (2,0) theory, so that is the perspective we will take. \\

We first consider the restriction of the M-theory C-field. If we consider M-theory in some 11-dim spacetime $Y$, then the C-field $\check C \in \check H^4(Y)$ is a degree-4 differential cocycle, with a flux quantization law shifted by $\frac 14 p_1(TY)$ \cite{Witten:1996md, Witten:1996hc}. That is, if we let $G$ denote the field strength of $\check C$ and let $a \in H^4(Y,\mathbb{Z})$ be its characteristic class, then we can write $G = G' + \frac 14 p_1(TY) $ for some $G' \in \Omega^4_{\mathbb{Z}}(Y)$. \\

We expect that the restriction of the M-theory C-field to $M$ should source the self-dual field strength $H$ of the B-field $\check B \in \check H^3(M)$ in the tensor multiplet of the 6d theory on the M5 brane worldvolume $M$. Crucially, this includes matching their characteristic classes up to a torsion class, so that it is not sufficient to simply write $dH = \check C_M$. However the M5 branes themselves are the magnetic branes sourced by the $\check C$, so that the periods of $G$ integrated around any 4-cycle linking the brane should be quantized. This means that $G$ diverges near the brane itself, and so $\check C$ is only defined on $Y\setminus M $.  \\

 How then can we restrict the C-field to $M$? What we can do is consider the restriction of the C-field to a 4-sphere bundle $\pi : \mathcal S\to M$, realized as the boundary of some tubular neighborhood of $M$ in $Y$. If we denote this as $\check C_{\mathcal S}$ then it was shown in \cite{Monnier_2015} that there is a well-defined restriction to $M$ defined by: 
 \begin{equation}
     \check C_M \equiv \frac 12 \pi_*(\check C_{\mathcal S}\cup \check C_{\mathcal S})
 \end{equation}
 where $\cup$ is the cup product on differential cohomology. However, this result relies crucially on the requirement that the Euler class $e(N)$ of the R-symmetry bundle must vanish. One way to see this is to simply consider the restriction of the characteristic class $a \in H^4(Y,\mathbb{Z})$ to some $a_M \in H^4(M,\mathbb{Z})$, as in \cite{Monnier_2015}. We can use the long exact sequence in cohomology associated the 4-sphere bundle:
 \begin{equation}
     0 \to H^4(M,\mathbb{Z}) \rightarrow H^4(\mathcal{S},\mathbb{Z}) \xrightarrow{\pi_*} H^0(M,\mathbb{Z}) \xrightarrow{e(N) \cup} \dots 
 \end{equation}
where the last map is given by taking the cup product with the Euler class. Therefore we see that in order to split this sequence so that $H^4(\mathcal S, \mathbb{Z}) \cong H^4(M,\mathbb{Z}) \oplus \mathbb{Z}$ we must require that the Euler class vanishes. Only then can we cleanly identify the characteristic class $a_M \in H^4(M,\mathbb{Z})$ of $C_M$ as the left component of this decomposition. \\

It is at this point that the Euler structure becomes prevalent. The specification of an Euler structure is necessary to having a well-defined restriction of the C-field to $M$. One might object to this, since the Euler class automatically trivializes in dimensions $d\leq 7$, as shown in \cite{Monnier_2015}. However, the definition of the anomaly theory from the descent procedure requires extension of the topological data to an 8-dim manifold, where the Euler class must be trivialized by hand by choice of an Euler structure. \\ 

A review of the definition of an Euler structure can be found in \cite{monnier2017anomalyfieldtheoriessixdimensional}. The idea is equivalent to the usual specification of any other type of tangential structure, such as an orientation or a spin structure. For example, recall that a spin structure of a rank-$n$ bundle $N$ is a choice of lift of the principal $SO(n)$-bundle to a $\text{Spin}(n)$-bundle. Here, a $\text{Spin}(n)$ bundle is defined by a classifying map into the classifying space $B\text{Spin}(n)$, which itself is defined as the homotopy fiber of $B\text{Spin}(n)\to BSO(n)\xrightarrow{f} K(\mathbb{Z}_2,2)$. The homotopy class of $[f]$ defines the Steifel-Whitney class $w_2(N) \in H^2(BSO(n), \mathbb{Z}_2)$. Therefore a choice of spin structure amounts to a choice of trivialization of $w_2(N)$. \\

Analogously, a choice of Euler structure on a rank-5 bundle $N$ amounts to a choice of trivialization of the Euler class $e(N)\in H^5(BSO(5),\mathbb{Z})$. Equivalently, it can be seen as a lift of the $SO(5)$ frame bundle to an $SO(5)[e]$ bundle. Here we define such a bundle using the classifying space $BSO(5)[e]$, which is in turn defined as the homotopy fiber of the fibration $BSO(5)[e]\to BSO(5) \to K(\mathbb{Z},5)$ which defines the Euler class $e\in H^5(BSO(5),\mathbb{Z})$.  \\

In our restriction of the C-field to $M$, the choice of Euler structure is a choice of top cohomology class $\beta$ on the fibers of the 4-sphere bundle $M$. The Euler structure can be thought of as a choice of decomposition of the degree-4 cohomology of $\mathcal{S}$ into ``fiberwise" and ``horizontal" components, as can be seen explicitly in the above long exact sequence. \\

For our purposes, it will be useful to define a differential cocycle refinement $\check \beta$ whose characteristic class is $\beta$, and to decompose the C-field as $\check C_{\mathcal{S}} = \check \beta + \pi^*(\check A)$, as in \cite{monnier2017anomalyfieldtheoriessixdimensional}, where $\check A_M$ is some differential cocycle on $M$ shifted by $w_4(TM\oplus N)$. Then the effective C-field on $M$ can be decomposed as:
\begin{equation}
    \check C_M = \check b + \check A_M  \ \ \ \ \ \ \ \  \text{where}  \ \ \ \ \ \ \ \ \check b \equiv \frac 12 \pi_*(\beta \cup \beta)
\end{equation}

This manifestation of the Euler structure features a key role in the presentation of the 7-dim anomaly theory. 

\subsection{Anomaly Field Theories of 6d $\mathcal{N}=(2,0)$ Theories}
\label{sec22}

We now review the results of \cite{monnier2017anomalyfieldtheoriessixdimensional}, which derives the 7-dim anomaly field theory for the anomaly of a 6d (2,0) theory with gauge group $A_n$. This derivation uses the M-theory realization of the theory on the worldvolume of a stack of $n+1$ M5 branes. \\

The idea of the computation is that we choose a 7-manifold $U$, which represents the bulk spacetime for the anomaly theory, and we assume the all the data specified in the previous section has been extended to $U$. The worldvolumes of the M5 branes are defined by $k$ sections of the $R$-symmetry bundle $N_U$. If we realize the M-theory background $Y$ as the total space of $N$, and consider the 4-sphere bundle $\tilde U \to U$ in $Y$ as enclosing the stack of M5 branes, then we can extend $\tilde U$ to a 4-sphere bundle $\tilde W \to W$, where $W$ is an 8-manifold satisfying $\partial W = U$. Finally we take the limit where the stack collapses so that there is no separation between the M5 branes. We then consider the M-theory 12-form Chern-Simons term:
\begin{equation}
    \text{CS} _{12}(\tilde W, \check C_{\tilde W, k}) = 2\pi i \int_{\tilde W} \frac 16 G \wedge G \wedge G - G\wedge I_8
\end{equation}
where $G$ is the field strength of the C-field $\check C_{\tilde W, k} = k\check \beta +\pi^*(\check A_W)$, and $I_8$ is:
\begin{equation}
    I_8 = \frac{1}{48} \Big[p_2(T\tilde W) - \Big(\frac{p_1(T\tilde W)}{2}\Big)^2 \Big] 
\end{equation}
The 7-dim anomaly theory for the 6d (2,0) theory is then obtained by integrating $CS_{12}$ over the fibers of $\tilde W$ to obtain an 8-form topological term for the anomaly on $W$, and then subsequently identifying the 7-dim anomaly theory in $U$ defined by this term. The final step is to identify and subtract out  the center of mass degrees of freedom of the stack of $k$ M5 branes. \\

The result of the computation is the following 7-dim anomaly theory for $A_n$ gauge group \cite{monnier2017anomalyfieldtheoriessixdimensional}:

\begin{subequations}\label{mon_anom}
    \begin{equation}
        \frak{An}_{A_n} = \Big( \mathcal{DF}_f^{\frac 12}\Big)^{\otimes (-n)} \otimes \Big( \mathcal{DF}^{\frac 14}_{\sigma} \Big)^{\otimes (-n)}  \otimes \frak{An}_{HWZ} \otimes \overline{\mathcal{WCS}}_G[A_n,0]
    \end{equation}
    \begin{equation}
         \frak{An}_{HWZ} \equiv  \Big(\mathcal{WCS}_P[\mathbb{Z}, -2\check b]\Big)^{\otimes \frac{n(n+1}{2}}\otimes \Big( \mathcal{BF}[-2\check b, \check C']\Big)^{\otimes \frac{n(n+1}{2}} \otimes \Big( \mathcal{CS}\frak{p}_2[\check b] \Big)^{\otimes \frac{(n+2)(n+1)n}{6}}
    \end{equation}
\end{subequations}
\\

We will briefly describe each factor in this theory. Further details can be found in \cite{monnier2017anomalyfieldtheoriessixdimensional}. 

\begin{enumerate}
    \item The term $\mathcal{DF}_f^{\frac{1}{2}}$ is a ``half Dai-Freed" theory, defined by the modified eta invariant $ \mathcal{DF}_f^{\frac{1}{2}} = \exp(i\pi \xi_f) $, which is related via the Atiyah-Patodi-Singer index theorem to the integral over the 8-form index density of the Dirac operator $D_{f,W}$:
    \begin{equation}
        \xi_f(U) = \int_W \Big[ \Big( -2I_8 + \frac{1}{12} p_2(N) + \frac 14 L(TW) \Big) - \text{index}(D_{f,W}) \Big]
    \end{equation}
    where $L(TW)$ is the degree-8 component of the Hirzebruch genus of $TW$. This is a ``half" Dai-Freed theory because it acts on spinors which obey a symplectic Majorana condition. 

    \item The term $\mathcal{DF}_{\sigma}^{\frac 14}(U) = \exp\Big( 2\pi i \frac 18 \eta_{\sigma}(U)\Big) $ is a ``quarter Dai-Freed" theory, expressed in terms of $\eta_{\sigma}(U) $, the eta invariant of the signature  of the Dirac operator on $U$. 

    \item The term $\mathcal{WCS}_P[\mathbb{Z}, -2\check b]$ is a prequantum Wu-Chern-Simons theory of $\check b$. Wu-Chern-Simons theories are higher-dimensional analogues of spin Chern-Simons theories \cite{Monnier_2017_2} which allow half-integer quantization at the expense of depending on a choice of Wu structure. The term ``prequantum" simply means that it is an invertible field theory whose partition function is just the exponentiated action, interpreted as an element of a Hermitian line associated to the boundary. 

    \item  The term $\mathcal{BF}[-2\check b, \check C]$ is the prequantum invertible theory defined by the 7-form cocycle associated to an 8-form BF-type term with action:
    \begin{equation}
         S_{BF} = \int_U [(-2\check b) \cup \check C]_{co}
    \end{equation}
    where the subscript indicates that we are treating the term in brackets as a differential cocycle of degree 8, and picking out the 7-form cocycle which defines it. 

    \item The term $\mathcal{CS}\frak{p}_2[\check b]$ is a Chern-Simons like prequantum invertible theory associated to the 8-form action:
    \begin{equation}
         S_{CSp_2}=\int_U \Big[ \frac 14 p_2(N) - 3\check b^2  \Big]_{co}
    \end{equation}

    \item The last term, $\overline{\mathcal{WCS}}_G[A_n,0]$ is the complex conjugate of a discretely gauged Wu Chern Simons theory. Unlike the other terms which are invertible, this gauged theory gives rise to the space of conformal blocks for the $A_n$ (2,0) SCFTs (see e.g. \cite{Gukov:2020btk}) . In this paper we will only be interested in the invertible parts of the anomaly field theory, and so will deal no more with this term.

\end{enumerate}

The discussion in \cite{monnier2017anomalyfieldtheoriessixdimensional} also conjectures a generalization of the anomaly theory \hyperref[mon_anom]{(2.6)} to any choice of gauge group $G$, which takes the same form but whose coefficients depend on the more general data of $\frak g$, such as the rank $r(\frak g)$, the dual Coxeter number $h_{\frak g}$, and the root lattice $\Lambda_{\frak{g}}$. See also \cite{Bonetti:2024etn} for a recent discussion of the anomalies for the D-type theories. 

\subsection{The Hopf-Wess-Zumino Term}
\label{sec23}

The main subject of this paper is to study anomaly matching of the 6d (2,0) theory between the superconformal fixed point at the origin and a point far out on the tensor branch. Here we review some results of this matching which are achieved with the ``Hopf-Wess-Zumino" term, as described in \cite{Intriligator_2000}. \\

The anomaly of the $\text{Spin}(5)_R$ symmetry of the 6d (2,0) theory has been known long before the full form \hyperref[mon_anom]{(2.6)}. The original anomaly can be expressed purely in terms of a background $SO(5)_R$-connection $A$ with field strength $F$, along with a gravitational piece expressed in terms of the curvature $R$. For gauge group $G$ the anomaly is characterized by the 8-form \cite{Witten:1996md}:
\begin{equation}\label{ken_anom}
     I_8(G) = r(G)\frac{1}{48}\Big[p_2(F)-p_2(R) + \frac{1}{4}(p_1(F)-p_1(R))^2\Big] + \frac{1}{24}c(G) p_2(F) 
\end{equation}

where $p_i$ are the Pontryagin classes and $r(G)$ is the rank of the gauge group. The coefficient $c(G)$ is called the 't Hooft anomly of the $SO(5)_R$ symmetry. For the case $G=SU(N)$ it is known to be $c(SU(N))=N^3-N$ \cite{Harvey_1998}. The quantity $c(G)$ also enters in the Weyl anomaly \cite{Henningson:1998gx}, and so can be interpreted as a $c$-function which should decrease along RG flows. \\ 

This defines the anomaly at the superconformal origin of the moduli space of supersymmetric vacua. The moduli space of the 6d theory with gauge group $G$ is $\mathcal M = (\mathbb{R}^5)^{r(G)} / W_G$ where $W_G$ is the Weyl group of $G$. If we move away from the origin then the $SO(5)_R$ symmetry is spontaneously broken, and we can consider a direction on the moduli space where it is broken to $SO(4)_R$. The effective theory then includes a $\sigma$-model parametrizing the configuration space $\phi : M \to SO(5)/SO(4)\cong \mathbb{S}^4$. \\

We can determine the anomaly of the theory at a point far out on the moduli space. The massless spectrum for the theory is that of a $\mathcal{N}=(2,0)$ theory associated with $H$ plus an additional $U(1)$ supermultiplet, where $H\times U(1)$ is the little group of the Cartan of $G$. The difference between the anomalies of the theory at the origin and the theory at this far-out point is:
\begin{equation}
    I_8(G) - I_8(H\times U(1) ) = \frac{1}{24} \Big( c(G) - c(H) \Big) p_2(F) 
\end{equation}
Since the full anomaly must be matched in the IR theory, then this difference must be made up for by a WZW term. Such a term was found in \cite{Intriligator_2000}, known as the ``Hopf-Wess-Zumino" term. It can be expressed by extending to a 7-dim bulk $\Sigma_7$, and if we assume $H^4(\Sigma_7)=0$ it takes the explicit form:
\begin{equation}\label{hwz}
    S_{WZ}  = \frac 16 (c(G)- c(H)) \int_{\Sigma_7} \Omega_3( \phi , A)\wedge d\Omega_3( \phi, A) 
\end{equation}
where $d\Omega_3(\phi,A) \equiv \frac{1}{2}\phi^*(e_4)$ is the pullback of the Euler class $e_4$ on $\mathbb{S}^4$. This term has the correct gauge variation to match the anomaly: $\delta \int_{\Sigma_7} \Omega_3\wedge d\Omega_3 = \frac14 \int_{M_6} \tilde p_2(A)$ where $\tilde p_2(A)$ is the 6-form derived from $p_2(F)$ by descent. The name ``Hopf-Wess-Zumino" derives from the fact that this term computes the $\pi_7$ winding of the $\phi$ configuration on $\mathbb{S}^4$ according the quaternionic Hopf fibration $\mathbb{S}^3\to \mathbb{S}^7\to \mathbb{S}^4$ when we set $A=0$. \\

The Hopf-Wess-Zumino term has since been studied further \cite{Fiorenza_2021, Kalkkinen_2003, Hu_2011, arvanitakis2019branewesszuminotermsaksz}. A qualitative explanation for this term is mentioned in \cite{Monnier_2014} in the framework of the computation of \hyperref[mon_anom]{(2.6)} for the $A_n$ series, arguing that this term originates from the fact that the bulk M-theory Chern-Simons term contributes an anomaly when evaluated on trapped spaces between the 4-sphere bundles over the separated M5 branes. This yields a difference with the anomaly calculated when the M5 branes are coincident. In \cite{Fiorenza_2021} the quantization of the Hopf-Wess-Zumino term is treated from the hypothesis that the proper Dirac quantization of the M-theory C-field uses a non-abelian generalized cohomology theory known as J-twisted cohomotopy. \\

Our goal in this paper is to provide an explanation for the origin of the Hopf-Wess-Zumino term which connects to the traditional understanding of anomaly matching with WZW terms, which we take up in \hyperref[sec4]{Section 4}.

\newpage

\section{The Anomaly from M5 Cobordism}
\label{sec3}

Now that we have an understanding of the anomaly theories for the 6d (2,0) theory, we can ask about their origin and classification (at least for the invertible parts). The common strategy for the classification of anomalies is by classifying the topological field theories which define them by inflow. The bulk anomaly field theory is classified as a topological invariant which can be evaluated on mapping class tori. For anomalies of ordinary invertible bosonic symmetries, the classification is by group cohomology, or equivalently classifying space cohomology. However it is believed that the most general classification of the deformation classes of anomaly theories are by invertible field theories defined by some choice of generalized cohomology \cite{Freed_2021, Kapustin_2015, Yonekura_2019, Gaiotto_2019, Xiong_2018, Freed_2006}. When the anomalous symmetry depends critically on some tangential structure such as orientation or spin structure, this generalized cohomology theory is defined as the Anderson dual to the bordism theory enriched with this tangential structure. The deformation classes of anomaly field theories are then associated to classes in this cohomology group, and include bordism invariants. A well-known example is in theories of fermions where there can exist a global contribution to the anomaly coming from the eta invariant, which is an invariant of spin bordism \cite{Dai:1994kq, Witten:2019bou, Garcia-Etxebarria:2018ajm}.  \\

Let us quickly review this modern characterization of anomalies. Suppose that we have a global symmetry $G$ which is anomalous. This anomaly is seen by coupling to a background $G$-gauge field, topologically classified by a map into the classifying space $BG$. When we attempt to define the bulk anomaly theory in $(d+1)$-dim which cancels the anomaly by inflow, then we can extend this map into the bulk. However for many anomalies this extension needs to be compatible with some tangential structure, such as the spin structure. The equivalence classes of $d$-dim manifolds which can be extended in this way while preserving both the spin structure and the map into $BG$ is the bordism group $\Omega_d^{spin}(BG)$. The deformation classes of invertible phases $\text{Inv}^{d+1}_{spin}(BG)$ which classify the topological class of the anomaly theory is then defined by the generalized cohomology $(D\Omega_{spin})^{d+2}(BG) $ known as the Anderson dual to this bordism, which fit into a short exact sequence:
\begin{equation}
    0 \to \text{Ext}_{\mathbb Z}(\Omega_{spin}^{d+1}(BG)) \to (D\Omega_{spin})^{d+2}(BG) \to \text{Hom}(\Omega^{d+2}_{spin}(BG),\mathbb Z)\to 0
\end{equation}
This combines the free pieces $\text{Hom}(\Omega^{d+2}_{spin}(BG),\mathbb Z)$ and the torsion pieces $\text{Ext}_{\mathbb Z}(\Omega_{spin}^{d+1}(BG)) = \text{Hom}(\text{Tors}(\Omega_{d+1}^{spin}(BG)) ,U(1))$ of the anomaly theory into a single topological invariant. The traditional anomaly polynomial which appears in the descent procedure derives from an element $\mathcal I_{d+2} \in \text{Hom}(\Omega^{d+2}_{spin}(BG),\mathbb Z) $, and is conventionally expressed in terms of characteristic classes in the background gauge fields and curvature through the relation $\text{Hom}(\Omega^{d+2}_{spin}(BG),\mathbb Z)\otimes \mathbb Q \cong H^{d+2}(B\text{Spin}\times BG, \mathbb Q)$ \cite{Lee_2021}. \\

It is then natural to ask whether the anomaly theory defined in \hyperref[mon_anom]{(2.6)} also fits into this classification. The strongest motivation for this question will really come in \hyperref[sec4]{Section 4}, when we attempt to understand the origin of the Hopf-Wess-Zumino term. There we will see that both the ordinary cohomology classification, and the classification by usual cobordism (orientation, spin, etc) fail to be sufficient. This failure seems reasonable given the list of exotic data we reviewed in \hyperref[sec21]{Section 2.1} which makes critical contributions to the overall anomaly theory \cite{monnier2017anomalyfieldtheoriessixdimensional, Monnier_2014}. In this paper we will argue that therefore a more exotic bordism theory which accounts for all of the 6d (2,0) structure is necessary to capture the full anomaly. 

\subsection{M5 Bordism}
\label{sec31}

Given that all the tangential structure we reviewed in \hyperref[sec21]{Section 2.1} makes a pivotal appearance in the anomaly \hyperref[mon_anom]{(2.6)}, we deem it reasonable to conjecture that the appropriate invertible phases which define this anomaly are classified by a cobordism theory which incorporates this structure. We will therefore attempt to construct such a bordism theory using the usual Pontryagin-Thom construction \cite{Freed_bordism}, and analyze the resulting invertible phases defined by the cobordism groups. \\

In the Pontryagin Thom construction, the bordism groups with a tangential structure $\mathcal{X}$ are defined by constructing a Thom spectrum $M\mathcal{X}$, and taking the stable homotopy groups, so that $\Omega^{\mathcal{X}}_k\equiv \pi^s_k(M\mathcal{X})$. If we want to enrich the bordism with a map into a space $Y$, then the Thom spectrum is augmented by a smash product with this space: $M\mathcal X\wedge Y_+$, so that $\Omega^{\mathcal X}_k(Y) = \pi^s_k(M\mathcal X\wedge Y_+) $. \\

Our goal is to construct a Thom spectrum which incorporates the following structure:

\begin{enumerate}
    \item The rank-5 $SO(5)[e]_R$ bundle $N$

    \item A spin structure on $TM \oplus N$

    \item An Euler structure on $N$

    \item The characteristic class $a_M \in H^4(M,\mathbb{Z}) $ of the C-field
    
\end{enumerate}

In fact such a bordism theory was proposed in \cite{monnier2017anomalyfieldtheoriessixdimensional}, referred to as ``M5 bordism". There the objective of this construction was not to study invertible field theories, but only to verify that 7-manifolds with this 6d (2,0) structure were null bordant up to torsion. Nevertheless, we can adopt this proposed definition of the M5 bordism, which we now briefly review. \\

In order to incorporate the $R$-symmetry bundle $N$, we need to enrich the bordism with a map into the classifying space $BSO(5)[e]$, which recall is defined as the homotopy fiber of $BSO(5)[e]\to BSO(5) \xrightarrow{e} K(\mathbb{Z}, 5)$, where the homotopy class of $e$ defines the Euler class. \\

Now in order to define the Thom spectrum for the spin and Euler structure we can use the usual procedure of the Pontryagin-Thom construction. This proceeds by considering the Thom space of the universal bundle $E\text{Spin} \times ESO(5)[e] \to B\text{Spin}\times BSO(5)[e]$ obtained from the pullback of $ E\text{Spin}\times ESO(5)\to B\text{Spin}\times BSO(5)$, and then taking the spectrification. However there is a subtlety here in the fact that we want the spin structure to be established on $TM \oplus N$, but we only want the Euler structure defined on $N$. The resolution is to define the Thom spectrum: $M\text{Spin}\wedge \Sigma^{-5} TBSO(5)[e]$, where here $M\text{Spin}$ is the normal Thom spectrum for spin structure, and $TBSO(5)[e]$ is the Thom space of the universal bundle $ESO(5)[e] \to BSO(5)[e]$. This can be seen as the correct construction by embedding $TM \oplus N$ into a sufficiently large sphere, and defining the normal bundle of $M$ to be $\nu$ and the normal bundle of $N$ to be $\nu'$. Then we have that $\nu \cong \nu' \oplus N$, and the condition that $TM \oplus N$ is spin is equivalent to the condition that $\nu'$ be spin. The space $\Sigma^{-5}TBSO(5)[e]$ then defines the Thom spectrum for this virtual vector bundle. For convenience of notation we will define $F_5\equiv TBSO(5)[e]$, so that the Thom spectrum is $M\text{Spin}\wedge \Sigma^{-5}F_5$. \\

Now the characteristic class $a_M\in H^4(M,\mathbb{Z})$ can be defined as the homotopy class of a map $[a_M] : M \to K(\mathbb{Z},4)$, so in order to incorporate this into our Thom spectrum we must enrich with a map into $K(\mathbb{Z},4)$. Therefore our total Thom spectrum is given by:
\begin{equation}
    M\text{Spin}\wedge \Sigma^{-5}F_5 \wedge K(\mathbb{Z},4)_+
\end{equation}
The desired M5 bordism groups are then defined as the stable homotopy groups of this Thom spectrum:
\begin{equation}\label{m5bordism}
    \Omega^{M5}_k = \pi_k^s(M\text{Spin}\wedge \Sigma^{-5}F_5 \wedge K(\mathbb{Z},4)_+) = \Omega^{spin}_{k+5}(F_5\wedge K(\mathbb{Z},4)
\end{equation}
The possible M5 invertible phases in dimension $d$ are then defined as the cobordism groups $(D\Omega^{M5})^{d+1}$, defined by the Anderson dual:
\begin{equation}\label{anddual}
     0 \to \text{Ext}_{\mathbb{Z}}(\Omega^{M5}_d,\mathbb{Z}) \to \text{Inv}_d^{M5} \to \text{Hom}(\Omega^{M5}_{d+1},\mathbb{Z}) \to 0
\end{equation}
Since M5 bordism captures all of the 6d (2,0) structure which appears in the 7d anomaly theory \hyperref[mon_anom]{(2.6)}, we propose that the phases $\text{Inv}_7^{M5}$ characterize the deformation classes of the invertible parts of this theory. We claim that these invertible phases should capture all invertible features of the anomaly \hyperref[mon_anom]{(2.6)}, as well as its generalizations to other choices of $\frak{g}$. We will see evidence for this in the next section.

\subsection{The 6d (2,0) M5 Invertible Phases}

In order to further study the anomaly theory from this perspective of M5 bordism, we need to evaluate the M5 bordism groups defined in \hyperref[m5bordism]{(3.3)}, as well as the associated Anderson dual groups \hyperref[anddual]{(3.4)}. We perform this calculation using the Adams spectral sequence:
\begin{equation}
     E_2^{s,t} = \text{Ext}_{\mathcal{A}}^{s,t}( H^*(M\text{Spin}\wedge \Sigma^{-5}F_5 \wedge K(\mathbb{Z},4)_+, \mathbb{Z}_2)) \\
    \Rightarrow \pi_{s-t}^s(M\text{Spin}\wedge \Sigma^{-5} F_5 \wedge K(\mathbb{Z},4)_+)_2
\end{equation}
We leave the details of this calculation to \hyperref[appa]{Appendix A}. Since we are interested in $\text{Inv}_7^{M5}$ then we calculate the bordism groups up to dimension 8. The results are listed in \hyperref[tab:m5groups]{Table 1}. \\

\begin{table}[]
    \centering
    \begin{tabular}{c|c}
    $k$ & $ \ \ \   \Omega^{M5}_k$ \ \ \   \\
    \hline
    
        0 & $\mathbb{Z}$  \\
        1 & 0 \\
        2 & 0 \\
        3 & 0 \\
        4 & $\mathbb{Z}^4$ \\
        5 & $\mathbb{Z}_2^3 $ \\
        6 & $\mathbb{Z}_2^5$ \\
        7 & 0 \\
        8 & $\mathbb{Z}^{11} $
    \end{tabular}
    \caption{M5 Bordism Groups}
    \label{tab:m5groups}
\end{table}

From this we find that $\text{Inv}_7^{M5}\cong \mathbb{Z}^{11}$ has 11 free generators.  In this work we do not attempt a thorough description of the precise topological invariants which correspond to each of these generators, however we can gather some immediate information about them using the details of the calculation in \hyperref[appa]{Appendix A}. In particular, we find that the generators are represented by $h_0$-towers in the $E_2$ page of the Adams spectral sequence. We note that these towers are in 1-1 correspondence with those generators of $H^8(M\text{Spin}\wedge \Sigma^{-5}F_5\wedge K(\mathbb{Z},4),\mathbb{Z}_2)$ which are annihilated by the first Steenrod square $Sq^1$. We list out these 11 cohomology classes as:
\begin{equation}\label{gens}
    W_2^4 \ \ , \ \ W_2^2W_4 \ \ , \ \ W_4^2 \ \ , \ \  w_4W_4 \ \ , \ \  w_4W_2^2 \ \ , \ \ w_8 \ \ , \ \ w_4^2 \ \ , \ \  W_4g \ \ , \ \ W_2^2g \ \ , \ \  w_4g \ \ , \ \ g^2  
\end{equation}
Here $W_i \in H^i(BSO(5)[e],\mathbb{Z}_2)$ correspond to the Stiefel-Whitney classes of the $BSO(5)[e]$ $R$-symmetry bundle. The $w_i \in H^i(M\text{Spin},\mathbb{Z}_2)$ are the Stiefel-Whitney classes of the tangent bundle coming from the spin structure contribution. Finally, $g$ is the generator of $H^4(K(\mathbb{Z},4), \mathbb{Z}_2) \cong \mathbb{Z}_2 $. We can then study these generators in order to gather clues about the nature of the generators of the invertible phases in $\text{Inv}_7^{M5}$. \\

Under our proposal, the 7d invertible phases which describe the anomaly field theory for the 6d (2,0) theory are constructed from the generators of the Anderson dual groups to these bordism groups, $\text{Inv}^{M5}_7=(D\Omega^{M5})^8\cong \mathbb{Z}^{11}$, whose generators can also be associated with $\hyperref[gens]{(3.6)}$. In this work we do not attempt to derive precise local expressions of these invertible phases, saving this precise matching for future work. However we can perform a quick comparison with the known anomaly theory \hyperref[mon_anom]{(2.6)} for the $A$-type theories. \\

In doing this, we see that all of the major features of \hyperref[mon_anom]{(2.6)} are present in our invertible phases. The generator $g \in H^4(K(\mathbb{Z},4), \mathbb{Z}_2) $ represents the presence of the $C$-field. The class $W_4 \in H^4(BSO(5)[e],\mathbb{Z}_2)$ represents the presence of the Euler structure, since the Euler structure trivializes $W_5\in H^5(BSO(5)[e],\mathbb{Z}_2)$, and this is what allows $Sq^1W_4=0$. The classes which capture the gravitational anomaly components resulting from the spin structure on the tangent and $R$-symmetry bundle are matched by the Dai-Freed-type theories in \hyperref[mon_anom]{(2.6)}. Terms like $W_4^2$ match the Wu-Chern-Simons terms for the Euler structure like $\mathcal{WCS}_P[\mathbb{Z}, -2\check b]$. Terms which couple the Euler structure and C-field like $W_4g$ are matched by the corresponding such term in \hyperref[mon_anom]{(2.6)} given by $\mathcal{BF}[-2\check b, \check C]$. \\

We therefore find encouraging evidence that M5 cobordism is the appropriate cohomology theory to classify our invertible phases. However the most promising results will come when we consider the matching of these anomalies on the tensor branch. There, we will argue that the M5 bordism invertible phases resolve the puzzle of the classification of the Hopf-Wess-Zumino term. \\

\section{Anomaly Matching and the Hopf-Wess-Zumino Term}
\label{sec4}

Now that we have an understanding of the correct bordism which defines the invertible part of our anomaly field theory, we turn our attention to the main issue of our concern: the matching of the anomaly on the tensor branch. We saw in \hyperref[sec23]{Section 2.3} that this was achieved in the original anomaly theory \hyperref[ken_anom]{(2.10)} by the Hopf-Wess-Zumino term \hyperref[hwz]{(2.12)}. However there are some puzzles regarding the appropriate origin and classification of this term. We begin by reviewing the traditional classifications of WZW terms, as well as their more modern characterization in terms of invertible phases. In \hyperref[sec42]{Section 4.2} we illustrate how these approaches fail to apply to the Hopf-Wess-Zumino term. \\

Our goal then is to find the correct classification of the Hopf-Wess-Zumino term. In doing this, we will also try to generalize the anomaly matching described in \hyperref[sec23]{Section 2.3} to the full anomaly theory \hyperref[mon_anom]{(2.6)}, which includes all the appropriate 6d (2,0) data. We will achieve this using the lessons we learned in \hyperref[sec3]{Section 3}, by proposing M5 bordism as the appropriate vehicle with which to classify our invertible phases. \\

\subsection{WZW Terms}
\label{sec41}

One of the most useful consequences of anomalies is that they are invariant under RG flows, and thus allow us to match some information about our theory between the UV and the IR. In the UV, we might have some degrees of freedom charged under an anomalous symmetry $G$ which contribute to the anomaly. However when we flow to the IR, in general some of these degrees of freedom might be integrated out, so that naively the theory does not seem to possess the same anomaly. In symmetry-breaking phases, it is the role of the WZW term to make up for this loss in order to reproduce the full anomaly in the IR. \\

In the case of spontaneous symmetry breaking $G\to H$, where the IR is described by a $\sigma$-model with target space the vacuum manifold $X=G/H$, then there is a construction which characterizes the topological class of the WZW term known as $\textit{transgression}$ \cite{Dijkgraaf1990, Witten1992, freed2007pionsgeneralizedcohomology, Lee_2021}. If $G$ is the anomalous symmetry in the UV, then traditionally the anomaly is classified by an anomaly polynomial $\alpha \in H^{d+2}(BG)$.  The WZW term, on the other hand, is described by a class $\beta\in H^{d+1}(X)=H^{d+1}(G/H)$. The specific relation between these classes is defined by a transgression map $\tau : H^{d+1}(X)\to H^{d+2}(BG)$. \\

Formally, the transgression appears as a differential in the Leray-Serre spectral sequence associated to the fibration $G/H\to BH \to BG$. This is a cohomology spectral sequence with second page:
\begin{equation}
    E_2^{p,q}= H^{p}(BG,H^q(G/H,\mathbb{Z}))\Rightarrow H^{p+q}(BH) 
\end{equation}
The groups in the cells in the leftmost column $E^{0,k}$ correspond to $H^k(G/H,\mathbb{Z})$, whereas the groups in the bottom row $E^{k,0}$ correspond to $H^k(BG,\mathbb{Z})$. In general, we will have classes in the bottom row belonging to $H^k(BG,\mathbb{Z})$ which do not survive to be reproduced in $H^k(BH,\mathbb{Z})$. It then often falls to differentials of the form $ d_k^{0,k-1} : E_k^{0,k-1}\to E_k^{k,0}$ to kill these classes. These differentials are our transgression maps $d_k^{0,k-1}\equiv \tau $. \\

However there is a more modern version of this story, which has developed from the new classifications for anomalies in terms of invertible phases defined by generalized cohomology. From this new perspective, the anomaly in the UV is classified as an invertible phase $\alpha \in \text{Inv}_{d+1}^{\mathcal{X}}(BG) \cong (D\Omega^{\mathcal X})^{d+2}(BG)$ defined by the Anderson dual of some bordism $\Omega^{\mathcal{X}}(BG) $ enriched with a $G$-symmetry. \\

Likewise, the modern understanding of WZW terms is that they are similarly classified by such invertible phases, except now enriched by a $\sigma$-model map into the target space $X$ \cite{freed2007pionsgeneralizedcohomology, Lee_2021, Freed_2018, saito2024wesszuminowittentermsspqcd, lee2021comments6dglobalgauge}. These are defined by invertible phases $\text{Inv}_{d+1}^{\mathcal{X}}(X)$, which are in turn defined by the short exact sequence:
\begin{equation}
    0 \to \text{Ext}_{\mathbb{Z}}(\Omega^{\mathcal{X}}_{d+1}(X),\mathbb{Z}) \to \text{Inv}^{\mathcal{X}}_{d+1}(X) \to \text{Hom}_{\mathbb{Z}}(\Omega^{\mathcal{X}}_{d+2}(X), \mathbb{Z}) \to 0 
\end{equation}
Our idea in the case of the generalized cohomology description of anomalies is to relate the topological classes of these invertible phases by transgression maps $ \tau : (D\Omega^{\mathcal{X}})^{d+1}(X) \to H^{d+2}(BG,\mathbb Z) $. Just as before where the transgression maps were defined as differentials of the cohomology LSSS, we define these new transgression maps by differentials of the \textit{generalized} Atiyah-Hirzebruch spectral sequence associated to the fibration $ G/H \to BH \to BG$, with the generalized cohomology theory being the Anderson dual of our chosen bordism. \footnote{A definition of transgression for differential $E^{n}$ generalized cohomology was constructed in \cite{freed2007pionsgeneralizedcohomology} in order to give an explicit description of the $SU(N)$ WZW terms. Here we seek only to detect the topological class of the WZW term, but it would be interesting to understand the relationship between this definition and the AHSS differential we use here.}\footnote{An interesting feature of this perspective on WZW terms is that they only transgress to features of the full anomaly in $(D\Omega^{\mathcal X})^{d+2}(BG)$ coming from the characteristic classes in $H^{d+2}(BG)$.}. This spectral sequence has second page:
\begin{equation}
    E_2^{p,q}=H^p(BG, (D\Omega^{\mathcal{X}})^q(X)) \Rightarrow (D\Omega^{\mathcal{X}})^{p+q}(BH) 
\end{equation}
The transgression maps arise as the differentials which stretch between the vertical and horizontal edges $\tau = d_k^{0,k-1} : E_k^{0,k-1} \to E_k^{k,0} $. Their role is to annihilate elements of $ H^d(BG,\mathbb{Z}) $ which do not survive to $(D\Omega^{\mathcal{X}})^k(BH)$. Classes $\beta \in E_k^{0,d+1}\cong (D\Omega^{\mathcal{X}})^{d+1}(X)$ which do not belong to the kernel of this map correspond to the invertible phases from which we get WZW terms. 

\subsection{Failure of the Naive Approach}
\label{sec42}

In the anomaly matching for the 6d (2,0) theory described in \hyperref[sec23]{Section 2.3}, the Hopf-Wess-Zumino term plays the role of a  WZW term for a low energy effective $\sigma$-model with target space $\mathbb{S}^4$. We can then try to use the above understanding of WZW terms to understand the origin of the HWZ term and its classification. \\ 

As a first attempt, we might try to use the traditional approach, and assume that both the anomaly and WZW term are characterized by cohomology classes. We immediately run into a problem with this approach. The cohomology classification implies that the WZW term in 6d should be characterized by an element of $H^7(\mathbb{S}^4,\mathbb{Z})$, which is clearly zero. This puzzle has been noted ever since the first proposal of the Hopf-Wess-Zumino term \cite{Intriligator_2000}. \\

We might then hope that the resolution lies in the more modern understanding of WZW terms as invertible phases classified by cobordism. We can attempt to calculate the usual bordism groups, which typically assume an orientation or spin structure. If we don't assume a spin structure, then the WZW term for the 6d theory would then be characterized as an element of $\text{Inv}^{SO}_6(\mathbb{S}^4)$, which is defined by the short exact sequence:

\begin{equation}
     0 \to \text{Ext}_{\mathbb{Z}}(\Omega^{SO}_{6}(\mathbb{S}^4),\mathbb{Z}) \to \text{Inv}^{SO}_{6}(\mathbb{S}^4) \to \text{Hom}_{\mathbb{Z}}(\Omega^{SO}_{7}(\mathbb{S}^4), \mathbb{Z}) \to 0 
\end{equation}

The relevant bordism groups can be easily evaluated. Note that we can always decompose $\Omega^{\mathcal{X}}_k(X)\cong \Omega^{\mathcal{X}}_k(pt)\oplus \tilde{\Omega}^{\mathcal{X}}_k $ where $\tilde{\Omega}^{\mathcal{X}}(X)$ denotes the reduced bordism. We can then use the suspension isomorphism for generalized cohomology by noting that $\mathbb{S}^4=\Sigma^4\mathbb{S}^0$ which tells us that $\tilde{\Omega}^{\mathcal{X}}_d(\mathbb{S}^4) \cong \Omega^{\mathcal{X}}_{d-4}(pt) $.\\

But then we note that $\Omega^{SO}_2(pt)=\Omega^{SO}_3(pt)=\Omega^{SO}_6(pt)=\Omega^{SO}_7(pt)=0$, so we conclude that $\text{Inv}^{SO}_6(\mathbb{S}^4)=0$, so there are no oriented invertible phases in 6d for the $\mathbb{S}^4$ $\sigma$-model, and therefore no WZW terms. \\

We might then think to try including a spin structure, to see if this helps to capture the WZW term. We can once again use the suspension isomorphism as above, and once again we find that $\Omega_3^{spin}(pt)=\Omega_6^{spin}(pt)=\Omega_7^{spin}(pt)=0$ while $\Omega_2^{spin}(pt)=\mathbb{Z}_2$. Therefore the spin invertible phases for the $\mathbb{S}^4$ $\sigma$-model are only classified by $\text{Inv}^{spin}_6(\mathbb{S}^4)=\mathbb{Z}_2$. This certainly is not sufficient to capture the WZW term, since the Hopf-Wess-Zumino term is known to have integer level quantization in order to match $c(G)-c(H)$ . \\

Thus we see that considering only the $R$-symmetry and orientation/spin structure is insufficient to capture the WZW term for the anomaly of the 6d (2,0) theory in the IR. Clearly something more is needed.

\subsection{Anomaly Matching with M5 Cobordism}
\label{sec43}

In order to resolve this puzzle, we can take inspiration from the lessons we learned about the anomaly in the UV. There we found that in order to capture all features of the anomaly, it was necessary to include all the additional M5 structure, including the total spin structure on $TM\oplus \mathcal{N}$, the Euler structure, and the characteristic class of the $C$-field. \\

Running with this idea, we can try to determine the invertible phases for the effective theory using M5 cobordism. However we must be careful about precisely how these invertible phases are defined. The M5 structure adds additional complications, so it is not so simple as evaluating the bordism groups enriched with a map into $\mathbb{S}^4$. We must think about specifically how the anomaly invertible field theory is defined with the UV structure and with the IR structure respectively. \\

Recall that our bordism theory in the UV was defined by:
\begin{equation}
    \Omega^{M5 \ (UV)}_k\equiv \Omega^{spin}_k(\Sigma^{-5} TBSO(5)[e]\wedge K(\mathbb{Z},4)) 
    = \Omega^{spin}_{k+5}(TBSO(5)[e]\wedge K(\mathbb{Z},4))
\end{equation}
Specifically, we note that $TBSO(5)[e]$ is the Thom space of the rank-5 $R$-symmetry bundle with classifying space $BSO(5)[e]$. However in the IR of our theory, the $R$-symmetry is spontaneously broken to $SO(4)$, so the $R$-symmetry bundle with Euler structure becomes a rank-4 bundle with classifying space $BSO(4)[e]$. The appropriate bordism group is then:
\begin{equation}
    \Omega^{M5 \ (IR)}_k\equiv \Omega^{spin}_k(\Sigma^{-4} TBSO(4)[e]\wedge K(\mathbb{Z},4)) 
    = \Omega^{spin}_{k+4}(TBSO(4)[e]\wedge K(\mathbb{Z},4))
\end{equation}
For tidiness, we define $\Gamma_{UV}\equiv \Sigma^{-5} TBSO(5)[e]\wedge K(\mathbb{Z},4) $ and $\Gamma_{IR} \equiv \Sigma^{-4} TBSO(4)[e]\wedge K(\mathbb{Z},4) $. In this notation then $\Omega^{M5 \ (UV)}_k = \Omega_k^{spin}(\Gamma_{UV})$ and $\Omega^{M5 \ (IR)}_k=\Omega^{spin}_k(\Gamma_{IR}) $.\\

Therefore, in order to understand the meaning of a WZW term in this context, we must retreat to our definition of a WZW term in terms of the transgression map. That is, we use the generalized Atiyah-Hirzebruch spectral sequence for the $D\Omega^{spin} $ cohomology theory associated to the fibration:
\begin{equation}
     X \to \Gamma_{IR} \to \Gamma_{UV}
\end{equation}
induced from the usual fibration $ \mathbb{S}^4\to BSO(4) \to BSO(5)$. Here the homotopy fiber $X$ is playing the role of our $\sigma$-model target space \footnote{A quick comment on the geometric interpretation of this abstract space $X$. Consider the Thom spectrum $\Sigma^{-5}TBSO(5)[e]$ which appears in $\Gamma_{UV}$. Notice that $TBSO(5)[e]$ is the Thom space of a rank-5 bundle, so by the Thom isomorphism its cohomology is equivalent to that of $BSO(5)[e]$, except that it is shifted up by 5 degrees. However taking $\Sigma^{-5}$ shifts the cohomology back down by 5 degrees, so in total the cohomology of $\Gamma_{UV}$ is isomorphic to that of $BSO(5)[e]\wedge K(\mathbb{Z},4)$. The only difference is that the action of the Steenrod squares on the cohomology has been twisted, so that the cohomology has a different Steenrod module structure. The same is true for $\Gamma_{IR}$. Therefore from a cohomology perspective $X$ ``looks" just like $BSO(5)[e]\wedge K(\mathbb{Z},4) / BSO(4)[e] \wedge K(\mathbb{Z},4)$, which ``looks" similar to $\mathbb{S}^4\wedge K(\mathbb{Z},4)$.}. \textit{It is the invertible phases enriched with maps into $X$ which will define our WZW terms}.\\

This spectral sequence has second page:
\begin{equation}
    E^{p,q}_2 = H^p( \Gamma_{UV}, (D\Omega^{spin})^q(X)) \Rightarrow (D\Omega^{spin})^{p+q}(\Gamma_{IR})
\end{equation}
The invertible phases which define our WZW term are elements of $\text{Inv}^{M5}_{k+1}(X)\equiv (D\Omega^{spin})^{k+2}(X) $. In our case, the invertible phases in the IR of the 6d (2,0) theory will be given by elements of $\text{Inv}^{M5}_{6}(X) $. Specifically, the WZW terms will be elements of $ \text{Inv}^{M5}_6(X) = (D\Omega^{spin})^7(X)$ which are sent by transgression maps $\tau : (D\Omega^{spin})^7(X) \to H^8( \Gamma_{UV}, \mathbb{Z}) $ to kill classes in the image. \\

Our objective is then clear: we should work backwards in the above spectral sequence from knowledge of $(D\Omega^{spin})^k(\Gamma_{IR}) $ and of the cohomology of $ \Gamma_{UV} $ in order to determine the groups $(D\Omega^{spin})^k(X) $ and their associated transgression maps. Elements of $(D\Omega^{spin})^7(X) $ which are mapped to nontrivial elements under the transgression map will then correspond to our WZW terms. \\ 

We leave the details of this analysis to \hyperref[appb]{Appendix B}, and simply summarize the results of our argument here. We list the invertible phases $\text{Inv}^{M5}_k(X) = (D\Omega^{spin})^{k+1}(X)$ through $k=6$ in \hyperref[table2]{Table 2}. \\

\begin{table}[]
    \centering
    \begin{tabular}{c|c}
    $k$ & $(D\Omega^{spin})^k(X)$ \\
    \hline
    
        0 & $\mathbb{Z}$  \\
        1 & 0 \\
        2 & $\mathbb{Z}_2$ \\
        3 & 0 \\
        4 & $\mathbb{Z}$ \\
        5 & 0 \\
        6 & $\mathbb{Z}_2^2\times \mathbb{Z} $ \\
        7 & $\mathbb{Z}_2\times \mathbb{Z} $ \\
    \end{tabular}
    \caption{Invertible Phases of Effective Theory}
    \label{table2}
\end{table}

We see that it indeed $\text{Inv}^{M5}_6(X) = (D\Omega^{spin})^7(X) = \mathbb{Z}_2\times \mathbb{Z} $ has a free factor. Moreover, we argue in the appendix that there is a transgression map $\tau = d_8^{0,7} : (D\Omega^{spin})^7(X)\to H^8(\Gamma_{UV}, \mathbb{Z}) $. Using the information about the 11 generators provided to us by the classes in \hyperref[gens]{(3.6)}, we find evidence that this transgression map  sends the generator of this free factor to a generator containing the 2nd Pontryagin class of the $R$-symmetry bundle $p_2 \in H^8(\Gamma_{UV}, \mathbb{Z})$. \\

Thus we find evidence of an invertible phase in $\text{Inv}^{M5}_6(X)$ with integer level quantization, which transgresses to $p_2$ of the $R$-symmetry. This is precisely what we expect of the Hopf-Wess-Zumino term, so we propose this phase as a natural candidate for the HWZ term in the IR effective theory. \\

We see that in order to discover the invertible phase corresponding to this WZW term, the inclusion of the M5 structure in our bordism theory was crucial. 

\section{Discussion}

In this work we have proposed that M5 cobordism, as defined in \cite{monnier2017anomalyfieldtheoriessixdimensional}, is the appropriate generalized cohomology theory with which to understand the invertible phases defining the anomaly theory of the 6d $\mathcal{N}=(2,0)$ SCFTs. We were led to this proposal by the appearance of additional structure in the full anomaly theory derived in \cite{monnier2017anomalyfieldtheoriessixdimensional}, including the Euler structure and C-field restriction. We successfully computed the cobordism groups through dimension 8, and found evidence that the associated invariants match the kinds of terms which appear in our anomaly theory. More compelling is the fact that this perspective on the anomaly clarifies the procedure of anomaly matching on the tensor branch when the R-symmetry undergoes SSB $SO(5)\to SO(4)$. There we found evidence for the existence of a WZW term which appears to play the role of the previously-derived Hopf-Wess-Zumino term. \\

This leaves open several avenues for future investigation in order to better understand and confirm this proposal. A better understanding of the 11 classes in $\text{Inv}_7^{M5}=(D\Omega^{M5})^8$ could be used to express the precise anomaly of the 6d (2,0) theories, and allow one to compare against the results of \cite{monnier2017anomalyfieldtheoriessixdimensional}. In turn, this would allow for the correspondence proposed in \hyperref[appb]{Appendix B} between these classes and the classes of the $E_2$ page in the symmetry-breaking spectral sequence to be placed on rigorous ground. This would then allow for a precise understanding of the transgression map, and a full proof that the transgression kills $p_2$.  It would also be useful to develop a more refined understanding of the precise role of this transgression map in the Atiyah-Hirzebruch spectral sequence for anomaly matching in the case of anomalies defined by generalized cohomology. \\

More fundamentally, it would be interesting to understand precisely how such a presentation of the 6d (2,0) anomaly in terms of M5 cobordism could be derived from M-theory directly. 

\newpage

\appendix

\section{Appendix A: M5 Bordism Calculations}
\label{appa}

From the discussion in \hyperref[sec3]{Section 3}, we would like to evaluate the bordism groups:
\begin{equation}
    \Omega^{M5}_d \equiv \Omega^{spin}_d( \Sigma^{-5}TBSO(5)[e]\wedge K(\mathbb{Z},4)) \\ \cong \Omega^{spin}_{d+5}(TBSO(5)[e]\wedge K(\mathbb{Z},4))
\end{equation}
for $d\leq 8$. By the Pontrjagin-Thom construction for bordism, we know that such bordism groups are equivalent to stable homotopy groups of a given Thom spectrum:
\begin{equation}
    \Omega^{spin}_d(X) \cong \pi_d^s(M\text{Spin}\wedge X_+)
\end{equation}
Our approach to the calculation will then be to evaluate these stable homotopy groups using the Adams spectral sequence. This is a spectral sequence which converges to the 2-primary parts of stable homotopy groups. It has second page:
\begin{equation}
    E_2^{s,t} = \text{Ext}_{\mathcal{A}}^{s,t}( H^*(M\text{Spin}\wedge X_+, \mathbb{Z}_2))\Rightarrow \pi_{s-t}^s(M\text{Spin}\wedge X_+)_2 
\end{equation}
where the subscript at the end indicates that this is just the 2-primary part, since we only consider the mod 2 Steenrod algebra $\mathcal{A}$. \\

Therefore we are interested in evaluating the spectral sequence: 
\begin{equation}
    E_2^{s,t} = \text{Ext}_{\mathcal{A}}^{s,t}( H^*(M\text{Spin}\wedge \Gamma_{UV}, \mathbb{Z}_2)) \\
    \Rightarrow \pi_{s-t}^s(M\text{Spin}\wedge \Gamma_{UV})_2 
\end{equation}
where for convenience we have defined $ \Gamma_{UV} \equiv \Sigma^{-5}TBSO(5)[e] \wedge K(\mathbb{Z},4)_+ $. This requires knowledge of the cohomology and Steenrod module structure of $M\text{Spin}\wedge \Sigma^{-5}TBSO(5)[e] \wedge K(\mathbb{Z},4)_+$. We will determine this information for each factor individually before combining them. \\

The Steenrod module structure of $M\text{Spin}$ is well-known. However since it will be useful later we explain explicitly how to determine it. Recall that the classifying space $B\text{Spin}$ is defined as the homotopy fiber for the fibration $B\text{Spin} \to BSO \to K(\mathbb{Z}_2,2) $ which defines the second Stiefel-Whitney class $w_2$. Therefore its mod 2 cohomology $H^*(B\text{Spin},\mathbb{Z}_2)$ can be determined from $H^*(BSO,\mathbb{Z}_2)$ by killing $w_2$ and all iterated Steenrod squares $Sq^{i_k}Sq^{i_{k-1}}\dots Sq^{i_1}w_2$. The resulting cohomology ring is then equivalent to that of $BSO$ with these relations enforced. Since this is a cohomology ring composed of (equivalence classes of) Stiefel-Whitney classes, then the action of the Steenrod squares can be determined by the Wu formula. 
\begin{equation}
    Sq^iw_j = \sum_{t=0}^i {j+t-i-1\choose t}w_{i-t}w_{j+t}
\end{equation}
Once we know the Steenrod algebra for $B\text{Spin}$ we can follow it through the definition of the Thom spectrum $M\text{Spin}$. This first requires taking the Thom spaces $TB\text{Spin}(k)$ of the universal bundles $E\text{Spin}(k)\to B\text{Spin}(k)$, and then taking the spectrification which involves taking the colimit as $k\to \infty$. The cohomology of the resulting space $M\text{Spin}$ can then be determined from the Thom isomorphism, which gives an isomorphism of cohomology $H^*(TB\text{Spin}(k),\mathbb{Z}_2)\cong H^{*+k}(B\text{Spin}(k),\mathbb{Z}_2)\cong H^*(B\text{Spin}(k),\mathbb{Z}_2)\cup \Phi_k$ where $\Phi_k$ is the Thom class. We then find that:
\begin{equation}
    H^*(M\text{Spin},\mathbb{Z}_2)\cong H^*(B\text{Spin},\mathbb{Z}_2)\cup \Phi
\end{equation}
where $\Phi$ is the universal Thom class. The action of the Steenrod squares can then be determined using the Wu formula and the fact that $Sq^i \Phi = w_i$ are the Stiefel-Whitney classes of the universal bundle. Computing all of this allows us to determine the Steenrod module structure of $H^*(M\text{Spin},\mathbb{Z}_2)$. \\

Next, let us turn out attention to the second factor $\Sigma^{-5} TBSO(5)[e]$. Recall from our definition of Euler structure that $BSO(5)[e]$ is defined as the homotopy fiber of the fibration $BSO(5)[e]\to BSO(5) \to K(\mathbb{Z},5)$ which defines the Euler class. Similar to $B\text{Spin}$, then the mod 2 cohomology $H^*(BSO(5)[e],\mathbb{Z}_2)$ can be determined from that of $BSO(5)$ by killing $w_5$ and all iterated Steenrod squares $Sq^{i_k}Sq^{i_{k-1}}\dots Sq^{i_1}w_5$. The cohomology of $BSO(n)$ is described in \cite{brown}. The action of the Steenrod squares on the resulting cohomology can then be determined by the Wu formula. \\

When we take the Thom space $TBSO(5)[e]$ then once again the Thom isomorphism relates the cohomology $H^*(TBSO(5)[e],\mathbb{Z}_2)\cong H^{*+5}(BSO(5)[e],\mathbb{Z}_2)\cong H^*(BSO(5)[e],\mathbb{Z}_2) \cup \Phi_5$ where $\Phi_5$ is the Thom class. Then taking $\Sigma^{-5}$ shifts the cohomology down by 5 again, so that we have $H^*(\Sigma^{-5}TBSO(5)[e],\mathbb{Z}_2)\cong H^*(BSO(5)[e],\mathbb{Z}_2)$. As in the case of $B\text{Spin}$ the action of the Steenrod squares is twisted by the contribution of $\Phi_5$, but can be determined by the Wu formula. \\

Finally, we turn to the last factor $K(\mathbb{Z},4)$, whose first few mod 2 cohomology groups and $\mathcal{A}_2$ Steenrod algebra is described in \cite{kz4}. The rest of the relevant Steenrod square actions can be determined by the Adem relations. \\

Once we have the cohomology and Steenrod actions for each element in $M\text{Spin}\wedge \Gamma_{UV}$, then we can determine the total cohomology and Steenrod algebra by repeated applications of the Kunneth formula. For our case the Kunneth formula simply reduces to:

\begin{equation}
    H^k(X\wedge Y, \mathbb{Z}_2) \cong \bigoplus_{i+j=k}H^i(X,\mathbb{Z}_2)\otimes H^j(Y,\mathbb{Z}_2)
\end{equation}

We calculate the full Steenrod module structure of $ H^k(M\text{Spin}\wedge \Gamma_{UV}, \mathbb{Z}_2) $ for $k\leq 9$. Once we have this in hand, it allows us to compute a projective resolution of the Steenrod module, whose cohomology can be evaluated to give the second page of the Adams spectral sequence $ E_2^{s,t} = \text{Ext}_{\mathcal{A}}^{s,t}( H^*(M\text{Spin}\wedge \Gamma_{UV}, \mathbb{Z}_2))$.  The diagram of our resulting second page is shown in \hyperref[adam]{Figure 1}

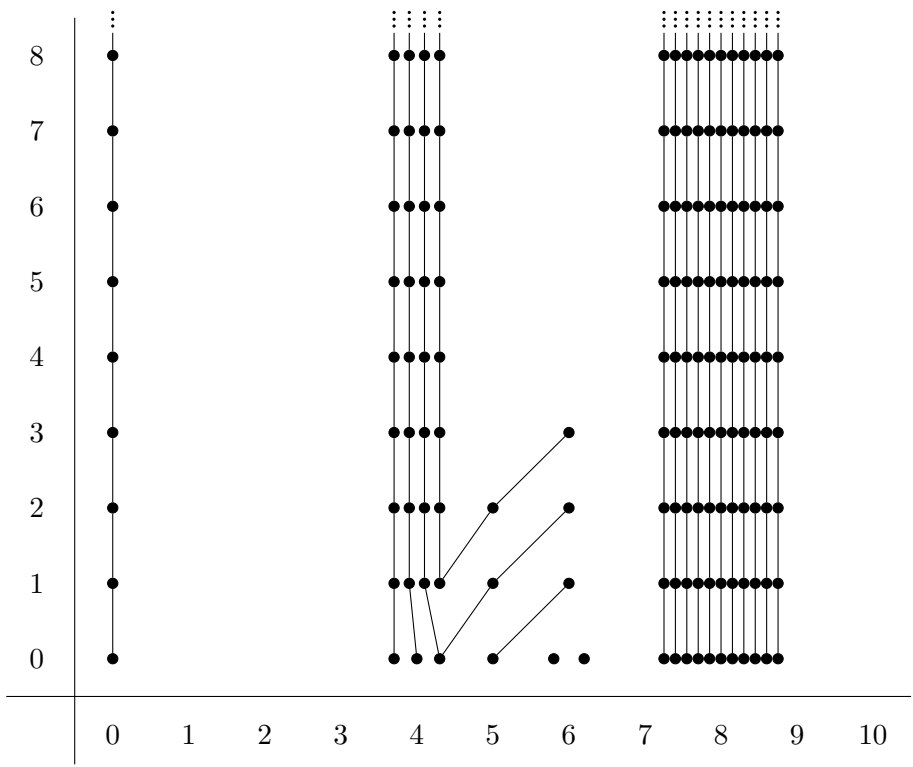
\begin{center}

\SseqNewClassPattern{line}{
(0,0);
(-0.2,0)(0.2,0);
(-0.3,0)(0,0)(0.3,0);
(-0.3,0)(-0.1,0)(0.1,0)(0.3,0);
(0,0)(0,0)(0,0)(0,0)(0,0);
(0,0)(0,0)(0,0)(0,0)(0,0)(0,0);
(0,0)(0,0)(0,0)(0,0)(0,0)(0,0)(0,0);
(0,0)(0,0)(0,0)(0,0)(0,0)(0,0)(0,0)(0,0);
(0,0)(0,0)(0,0)(0,0)(0,0)(0,0)(0,0)(0,0)(0,0);
(0,0)(0,0)(0,0)(0,0)(0,0)(0,0)(0,0)(0,0)(0,0)(0,0);
(-0.75,0)(-0.6,0)(-0.45,0)(-0.3,0)(-0.15,0)(0,0)(0.15,0)(0.3,0)(0.45,0)(0.6,0)(0.75,0);
}

\DeclareSseqGroup\tower {} {
\class(0,0)
\foreach \i in {1,...,11} {
\class(0,\i)
\structline{((0,i-1)(0,i)}
}
}

\begin{sseqpage}[class pattern=line, classes={circle,fill},degree = {-1}{1}, yscale = 1,
x range = {0}{10}, y range = {0}{8} ]\label{adam}
\tower(0,0)
\tower(4,0)
\tower(4,0)
\tower(4,0)
\tower(4,1)
\class(5,0)
\class(5,1)
\class(5,2)
\structline(4,0,3)(5,1)
\structline(4,1,4)(5,2)
\class(6,0)
\class(6,0)
\class(6,1)
\class(6,2)
\class(6,3)
\structline(5,0)(6,1)
\structline(5,1)(6,2)
\structline(5,2)(6,3)
\tower(8,0)
\tower(8,0)
\tower(8,0)
\tower(8,0)
\tower(8,0)
\tower(8,0)
\tower(8,0)
\tower(8,0)
\tower(8,0)
\tower(8,0)
\tower(8,0)

\end{sseqpage}
\end{center}

There are no Adams differentials in the relevant range. This allows us to read off the stable homotopy groups, and thus the M5 bordism groups. We list them in \hyperref[table3]{Table 3}. We note in particular that the 7th bordism group is trivial, as conjectured in \cite{monnier2017anomalyfieldtheoriessixdimensional}.  \\

\begin{table}[]
    \centering
    \begin{tabular}{c|c}
    $k$ & $\Omega^{M5}_k$ \\
    \hline
    
        0 & $\mathbb{Z}$  \\
        1 & 0 \\
        2 & 0 \\
        3 & 0 \\
        4 & $\mathbb{Z}^4$ \\
        5 & $\mathbb{Z}_2^3 $ \\
        6 & $\mathbb{Z}_2^5$ \\
        7 & 0 \\
        8 & $\mathbb{Z}^{11} $
    \end{tabular}
    \caption{M5 Bordism Groups}
    \label{table3}
\end{table}

This allows us to determine that the possible invertible phases $\text{Inv}^{M5}_7= (D\Omega^{M5})^8 \cong \mathbb{Z}^{11} $ have 11 free generators from which we can construct the anomaly field theory for the 6d $(2,0)$ theory. They correspond the the 11 $h_0$-towers in the $8^{th}$ column of the Adams $E_2$ page. \\

These $h_0$-towers are in 1-1 correspondence with the classes $\alpha \in H^8(M\text{Spin}\wedge \Gamma_{UV},\mathbb{Z}_2)$ which are annihilated by the first Steenrod square: $Sq^1\alpha = 0$. We list out these 11 classes:
\begin{equation}\label{gens}
    W_2^4 \ \ , \ \ W_2^2W_4 \ \ , \ \ W_4^2 \ \ , \ \  w_4W_4 \ \ , \ \  w_4W_2^2 \ \ , \ \ w_8 \ \ , \ \ w_4^2 \ \ , \ \  W_4g \ \ , \ \ W_2^2g \ \ , \ \  w_4g \ \ , \ \ g^2  
\end{equation}
Here $W_i \in H^i(BSO(5)[e],\mathbb{Z}_2)$ correspond to the Stiefel-Whitney classes of the $BSO(5)[e]$ $R$-symmetry bundle. The $w_i \in H^i(M\text{Spin},\mathbb{Z}_2)$ are the Stiefel-Whitney classes of the tangent bundle coming from the spin structure contribution. Finally, $g$ is the generator of $H^4(K(\mathbb{Z},4), \mathbb{Z}_2) \cong \mathbb{Z}_2 $.

\section{Appendix B: Anomaly Matching Spectral Sequence}
\label{appb}

From the discussion in \hyperref[sec3]{Section 3}, we are interested in the SSB fibration $X \to \Gamma_{IR} \to \Gamma_{UV}$ where we have defined $\Gamma_{UV}\equiv \Sigma^{-5}TBSO(5)[e]\wedge K(\mathbb{Z},4)$ and $\Gamma_{IR}\equiv \Sigma^{-4}TBSO(4)[e]\wedge K(\mathbb{Z},4)$. We would like to determine the invertible phases $\text{Inv}^{M5}_d(X)\equiv (D\Omega^{spin})^{d+1}(X)$ for $d\leq 6$. Our strategy to achieve this is to study the generalized Atiyah-Hirzebruch spectral sequence associated to the above fibration for the cohomology theory $D\Omega^{spin}$ . This spectral sequence has second page:
\begin{equation}
    E_2^{p,q} = H^p(\Gamma_{UV}, (D\Omega^{spin})^q(X)) \Rightarrow (D\Omega^{spin})^{p+q}(\Gamma_{IR})
\end{equation}
Our plan is to work backwards from our knowledge of $(D\Omega^{spin})^k(\Gamma_{IR})$ and of the cohomology of $\Gamma_{UV}$ in order to determine the invertible phases of the effective theory $(D\Omega^{spin})^{d+1}(X)$ associated to the fiber $X$. \\

First we must determine the groups $(D\Omega^{spin})^k(\Gamma_{IR})$. At first, it appears that we must redo the entire analysis of \hyperref[appa]{Appendix A} in order to evaluate the Adams spectral sequence for the space $\Gamma_{IR}$. However we can recall that all this spectral sequence depends on is the Steenrod module structure of $H^*(\Gamma_{IR},\mathbb{Z}_2)$. If we compare $\Gamma_{UV}\equiv \Sigma^{-5}TBSO(5)[e]\wedge K(\mathbb{Z},4)$ and $\Gamma_{IR}\equiv \Sigma^{-4}TBSO(4)[e]\wedge K(\mathbb{Z},4)$, then we see that they only differ in the first factor. Therefore we only need to analyze the differences in mod 2 cohomology between $\Sigma^{-5}TBSO(5)[e]$ and $\Sigma^{-4}TBSO(4)[e]$. \\

Recall that the way we determined the cohomology of this first factor in the case of $\Gamma_{UV}$ was first by killing all iterated Steenrod squares of the mod 2 reduction of the Euler class $w_5$, then taking the Thom space which shifted the cohomology up by 5, and then subsequently taking $\Sigma^{-5}$ which shifted it back down by 5. But notice that in $\Gamma_{IR}$ then $BSO(4)[e]$ automatically has vanishing $w_5$, as can be seen in \cite{brown}. Moreover taking the Thom space $TBSO(4)[e]$ shifts the cohomology up by only 4, so subsequently taking $\Sigma^{-4}$ takes us back to our original order. Moreover the twisting of the Steenrod action by the Thom class is equivalent to the case of $TBSO(5)[e]$ in $\Gamma_{UV}$. \\

From this, we conclude that in fact $\Gamma_{UV}$ and $\Gamma_{IR}$ have precisely the same mod 2 cohomology and Steenrod module structure, and therefore their bordism groups are equivalent, so $(D\Omega^{spin})^k(\Gamma_{IR}) = (D\Omega^{spin})^k(\Gamma_{UV})$. Thus we can just use the results of \hyperref[appa]{Appendix A}. We reproduce them here for convenience in \hyperref[tab4]{Table 4}. \\

\begin{table}[]
    \centering
    \begin{tabular}{c|c}
    $k$ & $(D\Omega^{spin})^k(\Gamma_{IR})$ \\
    \hline
    
        0 & $\mathbb{Z}$  \\
        1 & 0 \\
        2 & 0 \\
        3 & 0 \\
        4 & $\mathbb{Z}^4$ \\
        5 & 0 \\
        6 & $\mathbb{Z}_2^3 $ \\
        7 & $\mathbb{Z}_2^5$ \\
        8 & $\mathbb{Z}^{11} $
    \end{tabular}
    \caption{IR Dual Bordism Groups}
    \label{tab4}
\end{table}

The other piece of information we need is the cohomology $H^*(\Gamma_{UV})$ for both integral and mod 2 cohomology. We determined the mod 2 cohomology in \hyperref[appa]{Appendix A}. The integral cohomology can be likewise determined by knowing the integral cohomology of each factor. The integral cohomology of $BSO(5)[e]$ as well as the integral homology of $K(\mathbb{Z},4)$ is described in \cite{monnier2017anomalyfieldtheoriessixdimensional}, and so the cohomology can be determined by the universal coefficient theorem. Then the total integral cohomology can be determined by the Kunneth theorem. We quote the results in \hyperref[tab5]{Table 5}. The generators $\delta W_i \in H^i(BSO(5),\mathbb{Z}$ are the integral images of the Stiefel-Whitney classes, and $(2G), H$ are classes which appear in the cohomology of the fiber to make up for classes killed by elements of $H^*( K(\mathbb{Z}, 4), \mathbb{Z}) $. \\ 

\begin{table}[]
    \centering
    \begin{tabular}{c|c|c}
    $k$ & $H^k(BSO(5)[e],\mathbb{Z}) $ & \text{Generator}\\
    \hline
    
        0 & $\mathbb{Z}$ & 1 \\
        1 & 0 & - \\
        2 & 0 & - \\
        3 & $\mathbb{Z}_2$ & $\delta W_2$\\
        4 & $\mathbb{Z}^2$ &  $p_1, 2G$ \\
        5 & 0 & - \\
        6 & $\mathbb{Z}_2$ & $(\delta W_2)^2$ \\ 
        7 & $\mathbb{Z}_2^3$ & $p_1\delta W_2 , \delta(W_2W_4), \delta H $ \\
        8 & $\mathbb{Z}^4$ & $p_2, p_1^2, (2G)p_1, 8G^2 $
    \end{tabular}
    \caption{Integral Cohomology of $BSO(5)[e]$}
    \label{tab5}
\end{table}

Having determined the integral cohomology of $\Gamma_{UV}$, we can begin to write down the second page of the AHSS $E_2^{p,q} = H^p(\Gamma_{UV}, (D\Omega^{spin})^q(X))$. So far, all we know is the bottom row, which is $E^{k,0} = H^k(\Gamma_{UV}, \mathbb{Z})$. 
\begin{center}
\begin{tikzpicture}
  \matrix (m) [matrix of math nodes,
    nodes in empty cells,nodes={minimum width=3ex,
    minimum height=0.5ex,outer sep=-0.5pt},
    column sep=1ex,row sep=1ex]{
                \\
          8     \\                  
          7     \\                  
          6     \\                  
          5     \\                  
          4     \\                   
          3     \\            
          2     \\         
          1      \\
          0     &  \Z  & 0 &  0  & \Z_2 & \Z^3 & 0 & \Z_2 & \Z_2^5 & \Z^7  \\
    \quad\strut &   0  &  1  &  2  & 3 & 4 & 5 & 6 & 7 & 8\strut \\};
  
\draw[thick] (-3.1,-3.7) -- (-3.1,3.5) ;
\draw[thick] (-3.7, -2.9) -- (4, -2.9);
\end{tikzpicture}
\end{center}

Our goal is to determine $(D\Omega^{spin})^{k}(X)$ which will be given by the cells of the leftmost column. We can achieve this by knowing that this spectral sequence must converge to the groups $(D\Omega^{spin})^k(\Gamma_{IR})$, which are listed in \hyperref[tab4]{Table 4}. \\

In order to match $(D\Omega^{spin})^1(\Gamma_{IR})=0$ then we must have $E^{0,1}=0$, which means that the whole first row is also 0. Now we also know that $(D\Omega^{spin})^2(\Gamma_{IR})=0$, so whatever ends up being in $E^{0,2}$ should not survive to the $E_{\infty}$ page. We need to match $(D\Omega^{spin})^3(\Gamma_{IR})=0$ but we see that $E^{3,0}=\mathbb{Z}_2$. Therefore we must have that $E^{0,2}=\mathbb{Z}_2$, with its generator $a_2$ being mapped by the differential: $d_3^{0,2}(a_2)=\delta W_2 \in E_{3}^{3,0} $. Thus our updated $E_2$ page is:

\begin{center}
\begin{tikzpicture}
  \matrix (m) [matrix of math nodes,
    nodes in empty cells,nodes={minimum width=3ex,
    minimum height=0.5ex,outer sep=-0.5pt},
    column sep=1ex,row sep=1ex]{
                \\
          8     \\                  
          7     \\                  
          6     \\                  
          5     \\                  
          4     \\                   
          3     \\            
          2     & \ZZ_2 & 0 & \Z_2 & \Z_2 & \Z_2^3 & \Z_2 & \Z_2^4  \\         
          1     & 0 & 0 & 0 & 0 & 0 & 0 & 0 & 0   \\
          0     &  \Z  & 0 &  0  & \ZZ_2 & \Z^3 & 0 & \Z_2 & \Z_2^5 & \Z^7  \\
    \quad\strut &   0  &  1  &  2  & 3 & 4 & 5 & 6 & 7 & 8\strut \\};
  
\draw[thick] (-3.4,-3.7) -- (-3.4,3.5) ;
\draw[thick] (-3.8, -3) -- (4, -3);
\end{tikzpicture}
\end{center}

By the derivation property of the differentials, this also implies that $d_3^{3,2}(W_3a_2)=(\delta W_2)^2\in E_3^{6,0}$ as well as $d_3^{4,2}(W_2^2a_2)=\delta W_2 p_1\in E_3^{7,0}$ and $d_3^{4,2}(W_4a_2)=\delta W_2 (2\tilde g)\in E_3^{7,0}$. Finally, all the generators in $E_3^{6,2}$ are also killed. \\

We do not want to kill any factors of $\mathbb{Z}$ in $E^{4,0} = \mathbb{Z}^3 $ so we conclude that $E^{0,3}=0$. However in order to match $(D\Omega^{spin})^4(\Gamma_{IR})=\mathbb{Z}^4$ then we need one more $\mathbb{Z}$, so we conclude that $E^{0,4}=\mathbb{Z}$, and this factor is preserved until the $E_{\infty}$ page so all differentials send it to 0. By studying the generators of $E_{4,0} = H^4(\Gamma_{UV},\mathbb{Z})$, and the fact that the free factors in $(D\Omega^{spin})^4(\Gamma_{IR}) = \mathbb{Z}^4 $ are in 1-1 correspondence with those classes $W_4, W_2^2, g, w_4 \in H^4(M\text{Spin} \wedge \Gamma_{IR}, \mathbb{Z}_2)$ which are annihilated by $Sq^1$, then we can see that this extra factor of $\mathbb{Z}$ that we introduce in $E^{0,4} = (D\Omega^{spin})^4(X)$ might naturally be associated with $w_4 \in H^4(M\text{Spin} \wedge \Gamma_{IR}, \mathbb{Z}_2)$, which usually gives the background bordism invariant associated with $\Omega^{spin}_4(pt)$. \\

To match $(D\Omega^{spin})^5(\Gamma_{IR})=0$ then we must have $E^{0,5}=0$. Our updated $E_2$ page is now:

\begin{center}
\begin{tikzpicture}
  \matrix (m) [matrix of math nodes,
    nodes in empty cells,nodes={minimum width=3ex,
    minimum height=0.5ex,outer sep=-0.5pt},
    column sep=1ex,row sep=1ex]{
                \\
          8     \\                  
          7     \\                  
          6     \\                  
          5     & 0 & 0 & 0 & 0 \\                  
          4     & \Z & 0 & 0 & \Z_2  & \Z^3 \\                   
          3     & 0 & 0  & 0  & 0 & 0  & 0 \\            
          2     & \Z_2 & 0 & \Z_2 & \Z_2 & \Z_2^3 & \Z_2 & \Z_2^4  \\         
          1     & 0 & 0 & 0 & 0 & 0 & 0 & 0 & 0   \\
          0     &  \Z  & 0 &  0  & \Z_2 & \Z^3 & 0 & \Z_2 & \Z_2^5 & \Z^7  \\
    \quad\strut &   0  &  1  &  2  & 3 & 4 & 5 & 6 & 7 & 8\strut \\};
  
\draw[thick] (-3.4,-3.7) -- (-3.4,3.5) ;
\draw[thick] (-3.8, -3) -- (4, -3);
\end{tikzpicture}
\end{center}

We now need to match $(D\Omega^{spin})^6(\Gamma_{IR})=\mathbb{Z}_2^3$, but recall that one generator in $E^{6,0}$ and two generators in $E^{4,2}$ have already been killed, leaving only a single factor of $\mathbb{Z}_2$ on this diagonal. Thus we must have $E^{0,6}=\mathbb{Z}_2^2$. \\

We need to match $(D\Omega^{spin})^7(\Gamma_{IR})=\mathbb{Z}_2^5$, but recall that 2 generators in $E^{7,0}$ have been killed. But we see that these are compensated for by the other two factors of $\mathbb{Z}_2$ on this diagonal. Therefore we do not need to add anything to $E^{0,7}$ to match $(D\Omega^{spin})^7(\Gamma_{IR})$ . We momentarily leave open the possibility that $E^{0,7}$ contains other factors.  Our page then has the form: 

\begin{center}
\begin{tikzpicture}
  \matrix (m) [matrix of math nodes,
    nodes in empty cells,nodes={minimum width=3ex,
    minimum height=0.5ex,outer sep=-0.5pt},
    column sep=1ex,row sep=1ex]{
                \\
          8     \\                  
          7     & ? & 0 \\                  
          6     & \Z_2^2 & 0 & \Z_2^2 \\                  
          5     & 0 & 0 & 0 & 0 \\                  
          4     & \Z & 0 & 0 & \Z_2  & \Z^3 \\                   
          3     & 0 & 0  & 0  & 0 & 0  & 0 \\            
          2     & \Z_2 & 0 & \Z_2 & \Z_2 & \Z_2^3 & \Z_2 & \Z_2^4  \\         
          1     & 0 & 0 & 0 & 0 & 0 & 0 & 0 & 0   \\
          0     &  \Z  & 0 &  0  & \Z_2 & \Z^3 & 0 & \Z_2 & \Z_2^5 & \Z^7  \\
    \quad\strut &   0  &  1  &  2  & 3 & 4 & 5 & 6 & 7 & 8\strut \\};
  
\draw[thick] (-3.4,-3.7) -- (-3.4,3.5) ;
\draw[thick] (-3.8, -3) -- (4, -3);
\end{tikzpicture}
\end{center}

Let us look at the present state of our requirement that we match $(D\Omega^{spin})^8(\Gamma_{IR})=\mathbb{Z}^{11}$. One thing we immediately see is that there should be no torsion elements on this diagonal. Since the elements in $E^{0,6}$ survive to $E_{\infty}$ then by the derivation property of the differentials we believe there will be no nontrivial differentials starting from $E^{2,6}$. Therefore we need two factors of $\mathbb{Z}_2$ in $E^{0,7}$ to kill the $\mathbb{Z}_2^2 \in E^{2,6}$. \\

The generators of $(D\Omega^{spin})^8(\Gamma_{IR})=\mathbb{Z}^{11}$ are in 1-1 correspondence with the classes in \hyperref[gens]{(A.8)} which are annihilated by $Sq^1$. Let us compare these classes with the generators in $E^{8,0}\cong H^8(\Gamma_{UV},\mathbb{Z})$. Note that in the spectral sequence calculation of $H^*(BSO(5)[e],\mathbb{Z})$ in \cite{monnier2017anomalyfieldtheoriessixdimensional} that $(2\tilde g)$ plays the role analogous to $W_4$. We can then observe the following correspondences between some of the generators of these two groups: 
\begin{equation}\label{assoc}
     p_1g \leftrightarrow W_2^2 g \ \ \ , \ \ \ (2\tilde g)g \leftrightarrow W_4g \ \  \ , \ \ \ p_1^2 \leftrightarrow W_2^4 \ \ \ , \ \ \ (2\tilde g)p_1\leftrightarrow W_4W_2^2 \ \ \ , \ \ \  8\tilde g^2 \leftrightarrow W_4^2 \ \ \ , \ \ \ g^2 \leftrightarrow g^2 
\end{equation}
We also see that the cell $E^{4,4}$ introduces 3 more generators $a_4p_1, a_4(2\tilde g), a_4 g$. We argued above that $a_4$ contributes the generator associated to $w_4$, so we propose that these 3 generators can be associated to $w_4W_2^2, w_4W_4, w_4g $. \\

However if we look at all the generators in $(D\Omega^{spin})^8(\Gamma_{IR}) $ that need to be matched, we note that there are no present generators in the current $E_2$ page which match the generators associated to $w_4^2$ and $w_8$. These classes are typically associated to the generators of $\Omega^{spin}_8(pt)\cong \mathbb{Z}^2$, and we saw earlier that the responsibility of matching $w_4$ fell to $E^{0,4}$. We therefore propose that $E^{0,8}$ must contain a $\mathbb{Z}^2$, and that the associated generators will have the duty of matching $w_4^2$ and $w_8$. \\

But now, we can count that there are 12 generators along the diagonal which converge to $(D\Omega^{spin})^8(\Gamma_{IR})\cong \mathbb{Z}^{11}$, whereas we only need to match 11 generators. The only possibility left to us is that $E^{0,7}$ contains a factor of $\mathbb{Z}$. If we call the generator of this factor $a_7$, then we must have that this generator is sent by the differential $d_8^{0,7}: E_8^{0,7}\to E_8^{8,0}$ to kill a generator in $E^{8,0}\cong H^8(\Gamma_{UV},\mathbb{Z})$, in order to give us the correct number of generators in the $E_{\infty}$ page. Thus, our completed $E_2$ page is:

\begin{center}
\begin{tikzpicture}
  \matrix (m) [matrix of math nodes,
    nodes in empty cells,nodes={minimum width=3ex,
    minimum height=0.5ex,outer sep=-0.5pt},
    column sep=1ex,row sep=1ex]{
                \\
          8     & \Z^2 \\                  
          7     & \ZZ_2^2\times \Z & 0 \\                  
          6     & \Z_2^2 & 0 & \ZZ_2^2 \\                  
          5     & 0 & 0 & 0 & 0 \\                  
          4     & \Z & 0 & 0 & \Z_2  & \Z^3 \\                   
          3     & 0 & 0  & 0  & 0 & 0  & 0 \\            
          2     & \Z_2 & 0 & \Z_2 & \Z_2 & \Z_2^3 & \Z_2 & \Z_2^4  \\         
          1     & 0 & 0 & 0 & 0 & 0 & 0 & 0 & 0   \\
          0     &  \Z  & 0 &  0  & \Z_2 & \Z^3 & 0 & \Z_2 & \Z_2^5 & \ZZ^7  \\
    \quad\strut &   0  &  1  &  2  & 3 & 4 & 5 & 6 & 7 & 8\strut \\};
  
\draw[thick] (-3.7,-4) -- (-3.7,3.5) ;
\draw[thick] (-4.3, -3.3) -- (4.5, -3.3);
\end{tikzpicture}
\end{center}

The final question we can ask is \textit{which} generator in $E^{8,0}\cong H^8(\Gamma_{UV},\mathbb{Z})$ does $a_7$ kill? If we trust the above natural associations in \hyperref[assoc]{(B.2)}, then the only candidate left to us is the second Pontryagin class $p_2$.  Therefore we propose that $d_8^{0,7}(a_7)=p_2$. \\

We are now free to read off the left column of our $E_2$ page, which gives us the invertible phases of our effective theory $\text{Inv}_k^{M5}(X)= (D\Omega^{spin})^{k+1}(X)$. We list them in \hyperref[tab6]{Table 6}. \\

We indeed see that $\text{Inv}_6^{M5}(X) = (D\Omega^{spin})^7(X) $ appears to contain a factor of $\mathbb{Z}$, and so may define a WZW term with integer quantization. Moreover, the generator of this group is mapped under $d_8^{0,7}$ to kill $p_2 \in H^8(\Gamma_{UV},\mathbb{Z})$. This differential $d_8^{0,7}$ is by definition a transgression map $\tau$. Therefore we see that the candidate WZW invertible phase appears to transgress to the second Pontryagin class $p_2$, just as we expect for the Hopf-Wess-Zumino term. 

\begin{table}[]
    \centering
    \begin{tabular}{c|c}
    $k$ & $(D\Omega^{spin})^k(X)$ \\
    \hline
    
        0 & $\mathbb{Z}$  \\
        1 & 0 \\
        2 & $\mathbb{Z}_2$ \\
        3 & 0 \\
        4 & $\mathbb{Z}$ \\
        5 & 0 \\
        6 & $\mathbb{Z}_2^2$ \\
        7 & $\mathbb{Z}_2^2\times \mathbb{Z}$ \\
        
    \end{tabular}
    \caption{Effective Theory Invertible Phases}
    \label{tab6}
\end{table}

\newpage

\acknowledgments

AS is supported by Simons Foundation award 568420 (Simons Investigator) and award 888994 (The Simons Collaboration on Global Categorical Symmetries). It is my pleasure to thank Ken Intriligator and John McGreevy for helpful discussions. Thanks also to Jordan Benson for advice on the spectral sequence calculations.


\bibliographystyle{JHEP}
\bibliography{biblio.bib}


\end{document}